# ReaxFF Reactive Force Field Development for Cu/Si Systems and application to Copper Cluster Formation During Cu Diffusion Inside Silicon

Kamyar Akbari Roshan[1], Mahdi Khajeh Talkhoncheh[2], Jonathan E. Mueller[3], William A. Goddard III[4], Adri C.T. van Duin[2,5*]

[1] Department of Electrical Engineering, Pennsylvania State University, University Park, Pennsylvania 16802, United States

[2] Department of Chemical Engineering, Pennsylvania State University, University Park, Pennsylvania 16802, United States

[3] Helmholtz Institute Ulm (HIU) Electrochemical Energy Storage, 89081 Ulm, Germany

[4] Caltech, Division of Chemistry and Chemical Engineering, 1200 E California Blvd, Pasadena, CA 91125, USA

[5] Department of Mechanical Engineering, Pennsylvania State University, University Park, Pennsylvania 16802, United States

[*] Corresponding author. Email: acv13@psu.edu (Adri C. T. van Duin)

## Abstract

Transition metal impurities such as nickel, copper, and iron, in solid-state materials like silicon have a significant impact on the electrical performance of integrated circuits and solar cells. To study the impact of copper impurities inside bulk silicon on the electrical properties of the material, one needs to understand the configurational space of copper atoms incorporated inside the silicon lattice. In this work, we performed ReaxFF reactive force field based molecular dynamics simulations, studying different configurations of individual and crystalline copper atoms inside bulk silicon by looking at the diffusional behavior of copper in silicon. The ReaxFF Cu/Si parameter set was developed by training against DFT data, including the energy barrier for an individual Cu-atom inside a silicon lattice. We found that the diffusion of copper atoms has a direct relationship with the temperature. Moreover, it is also shown that individual copper atoms start to clusterize inside bulk silicon at elevated temperatures. Our simulation results provide a comprehensive picture of the effects of temperature and copper concentration on the crystallization of individual copper inside silicon lattice. Finally, the stress-strain relationship of $Cu_/Si$ compounds under uniaxial tensile loading have been obtained. Our results indicate a decrease in the elastic modulus with increasing level of Cu-impurity concentration. We observe spontaneous microcracking of the Si during the stress-strain tests as a consequence of the formation of a small Cu clusters adjacent to the Si surface.

* Correspondence should be addressed to acv13@engr.psu.edu

# 1  Introduction

The control of impurity contamination in the manufacturing of semiconductor devices has been identified as a critical issue due to their severe impact on the performance of microelectronics [1]. Metallic impurities can be introduced during the crystal formation or the fabrication process[2–4]. It has been reported that the 3d transition metal impurities like Mn, Fe, Co, Ni, and Cu in silica and silicon compounds introduce a variety of chemical activities contributing to the alterations in device performance[5]. Nowadays, due to copper's lower bulk resistivity and high activation energy compared to aluminum, copper is considered as a suitable alternative for Al in the design of ultra-large-scale integration devices [6]. However, copper has a faster diffusion mechanism in silicon/silica structures in comparison to the other 3d metals making it capable of quickly scattering over a silicon structure only in a few hours which results in reliability issues[7]. Besides, copper has a high interstitial solubility at high temperatures[8]. The combination of high diffusivity and having a steeply decaying solubility with temperature generates a high driving force for copper precipitation upon cooling. The fast diffusion and clustering of copper ions in dielectric materials play a crucial role in optimizing the performance of microelectronics[9–12], photonics [9,13], and electrochemical metallization cells[14] . The copper concentrations in bulk silicon crystals are negligible, as previously reported for iron and nickel[15,16]. However, due to its higher electronegativity compared to that of silicon[16,17], copper can quickly be produced on clean silicon wafer surfaces during any wet chemical process[18–20]. Therefore, it is critical to minimize the copper impurity content in chemicals used in device-production lines. Obtaining a more profound knowledge of the underlying molecular rearrangements that take place during the diffusion process is an important step to understand the evolution of cluster structures. This knowledge can eventually help us to overcome the disadvantage of copper contaminations for interconnected technology.

The molecular dynamics (MD) simulation technique is a relevant method for evaluating mechanical properties , the chemistry of Cu diffusion and its dissolution inside silicon structure. Among various MD methods, the ReaxFF reactive force fields (ReaxFF) [21,22] can be a very useful tool to gain information about interatomic interaction model which is available for a wide range of materials including organic-inorganic[23–30], organic-metal[31–33] and metal-metal [34,35] hybrid materials. Particularly, ReaxFF has been utilized broadly to investigate complicated combustion phenomena[36–38], interphase chemistry[39], and mechanical properties of various materials [40] and catalysts[41] . Recently, several systems, including both silica compounds and metals, have been investigated [5,42,43]. Urata et al. studied the interaction between metal copper and amorphous silica (a-$SiO_2$) as a primitive metal/oxide interaction by using ReaxFF to understand the intrinsic role of an oxidized metal layer on adhesion between a metal and an oxide glass [44] . In another work, the research carried out by Shirai et al. simulated the fast diffusion of Cu by first-principles calculations[7]. The simulations have demonstrated a clear migration of copper atoms between adjacent cells, and the diffusion constants agree with the experimental data using the Arrhenius equation. Nonoda et al. investigated the stability of Fe, Cu, and Ni atoms gettering in the large-scale integrated process near the (001) Si surface by performing DFT calculations[43]. They obtained the formation energy of Fe, Cu, and Ni atoms at the interstitial tetrahedral (T)-site in each atomic layer of Si super cell. It is worth to mention that during Cu diffusion, crystalline silicon (c-Si) undergoes a volumetric expansion, leading to crack nucleation and growth. Cu cluster formation in various sizes is an energetically favorable process which happens during Cu diffusion inside Si lattice as compared to isolated ions[14]. This indicates that the formation of metallic clusters does not require overcoming a nucleation barrier if the process involves long timescales that allow the silica atoms to relax around the Cu cluster. Moreover, the stability of a copper cluster in silicate and the underlying mechanism of the reaction of a switching cell on a copper electrode has been studied by Guzman et al[14]. Recent TEM studies have revealed the atomistic mechanism and dynamics of Cu diffusion inside c-Si[6]. The stress analysis of metals transport in Si lattice during diffusion processes has also been a subject of research. Ostadhossein et al. carry out ReaxFF-based MD simulations to examine lithium diffusion dynamics in silicon nanowires at the atomic scale during the lithiation process[40]. Their stress analysis has been presented that lithiation induces compressive stress, causing retardation or even the stagnation of the reaction front, also in good agreement with TEM observations.

Although there are works studying the fast diffusion of copper in silicon[7], the underlying mechanism of its rapid diffusion remains unclear. In addition, there are limited amounts of studies investigating copper clustering mechanism inside silicon. In the present work, we developed the Cu/Si force field parameters to understand the mechanism of the specific combination of the functional groups and molecular structures which gives us the best selection for describing Cu and Si atoms interactions during Cu diffusion and clustering. We compared the migration barriers of Cu in a-Si obtained by ReaxFF with those obtained by DFT calculations to validate the Cu/Si ReaxFF force field. In addition, the mechanical properties of the Si lattice after Cu diffusion have been investigated. Our results demonstrate the decisive role of stresses in the migration of the Cu inside pristine Si, which leads to micro cracks propagation. Lastly, we studied various configurations copper atoms incorporate inside the silicon lattice. We aimed to elaborate a new computational study to atomistically model different arrangements of individual copper atoms in bulk silicon under various conditions such as temperature and the level of Cu contamination.

Utilizing reactive force field enables us to simulate bond formation/breakage along with the dynamics of large molecular systems. In this work, we studied the dependence of Cu crystallization on the contamination level as well as environmental conditions such as temperature.

## 2 Methods

### 2.1 Atomistic Scale ReaxFF Simulations

The ReaxFF reactive force field method is a bond order (BO) based empirical potential that accounts for reactions by allowing bond formation and bond breakage during molecular dynamics simulation of chemical reactive systems [45,46]. Similar to nonreactive MD models, ReaxFF consists of two sets of terms: bonded and nonbonded (van der Waals and electrostatic interactions). However, ReaxFF allows bond formation and dissociation, and hence has significantly different bonded terms compared to those in classical potentials. ReaxFF uses a bond order-bond distance relation [36] in conjunction with the bond order-bond energy relation, which enables it to properly simulate the smooth formation and dissociation of bonds. All of the connectivity-dependent terms such as bond, angle, and torsion terms are made bond order dependent so that their contribution will diminish if the bond breaks. However, nonbonded interactions such as van der Waals and Coulomb are calculated between every pair of atoms irrespective of their connectivity. Though the nonbonded interactions are not bond order dependent, they are highly dependent on the distance of the atom pairs. Therefore, these contributions need to be updated at each step of the simulation. ReaxFF calculates atomic charges by using the electronegativity equalization method[47]. Additionally, to eliminate discontinuities in the nonbonded interaction energies and to reduce the range of the Coulomb interactions, a seventh order Taper function is employed[48,49]. The Taper function ensures that all nonbonded terms, together with their first, second, and third derivatives, go to zero at the nonbonded cutoff distance, which is typically picked to be 10 Å[50]. In short,

ReaxFF uses the following equation to find the energy of the system:

$$E_{system} = E_{bond} + E_{over} + E_{under} + E_{lp} + E_{val} + E_{tor} + E_{vdWaals} + E_{Coulomb} + E_{trip} \quad (1)$$

Where $E_{bond}$, $E_{over}$, $E_{under}$, $E_{lp}$, $E_{val}$, $E_{tor}$, $E_{vdWaals}$, $E_{Coulomb}$, and $E_{trip}$ represent bond energy, over-coordination energy penalty, under-coordination stability, lone-pair energy, valence angle energy, torsion angle energy, van der Waals energy, Coulomb, and triple bond stabilization energies, respectively. A more detailed description can be found in previous ReaxFF-related articles[51–54] Combined with energetic and conformational data from considerable QM calculations, ReaxFF can provide atomistic descriptions of many complex chemical reactions and can be adopted to simulate the interaction between Cu and Si atoms.

## 2.2 Force Field Development

The quality of a molecular dynamics simulation depends on the accuracy of the force field parameters; therefore, these parameters need to be trained against available experimental or quantum mechanical-based density functional theory (DFT) data[55,56]. In order to initiate the simulation on a relatively large-scale system to study the chemistry of Cu diffusion barrier inside Si system, we have developed a Si/Cu reactive force field parameter sets to include the descriptions of Cu and Si atoms interactions during dissolution and clustering. These parameters have been trained against DFT data including equation of state (EOS), Cu Diffusion data, and heats of formation of crystalline phases and molecules. The optimization of the parameters was performed via a successive one-parameter search technique to minimize the sum of error values:

$$Error = \sum_{i}^{n} [{}^{x_{i,QM} - x_{i,Reaxff}}/{\sigma_i}] \quad (2)$$

where $x_{i,QM}$ is the QM value, $x_{i,Reaxff}$ is the ReaxFF calculated value, and $\sigma_i$ is the weight assigned to data point, $i$.

To parameterize the ReaxFF EOS data, we carried out QM calculations for the EOS in various molecular species, such as $CuSi_8$, $Cu_2Si_8$, $Cu_3Si_8$, and $Cu_4Si_8$. **Fig. 1** compares ReaxFF and QM results for Birch–Murnaghan isothermal equations of state (a-e), and copper diffusion barrier (f). In each case, we constructed ground state geometries through full geometry optimization. Birch–Murnaghan equation of state calculation[57] was performed on crystalline $Cu_xSi_y$ (various space groups). We carried out periodic QM calculation based on density-functional theory (DFT). The Vienna ab initio simulation package (VASP) was used to solve the Kohn-Sham equations with periodic boundary conditions and a plane-wave basis set[58,59]. We applied Blöchl's all-electron frozen core projector augmented wave (PAW) method [60], and electron exchange and correlation is treated within the generalized gradient approximation (GGA) of PBE [61]. We applied compression and expansion with respect to the equilibrium volume of the crystal to calculate QM energies at different volume state. Next, during force field optimization, the corresponding energy values to each volume calculated by ReaxFF are compared with the QM data. **Fig. 1(a-d)** shows the EOS of the $Cu_xSi_y$ crystal as predicted by ReaxFF and QM.

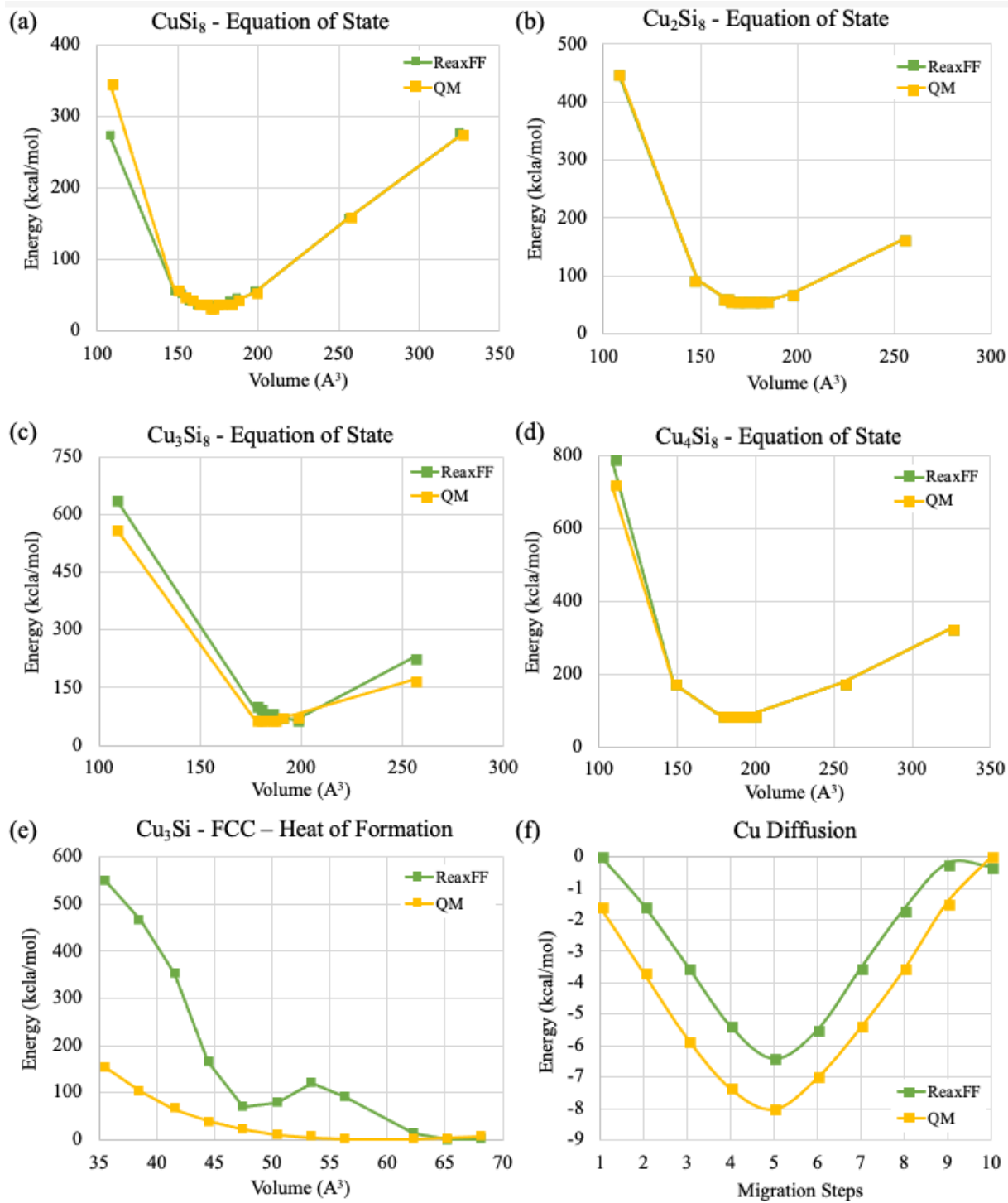

**Figure 1.** (a-d) Birch–Murnaghan equation of state for four different $Cu_xSi_8$ species ($CuSi_8$, $Cu_2Si_8$, $Cu_3Si_8$, $Cu_4Si_8$), using QM and ReaxFF methods. (e) Birch–Murnaghan equation of state for Fcc-$Cu_3Si$ using QM and ReaxFF methods. (f) Copper migration energy barrier to nearby adjacent vacancy site in silicon lattice calculated by ReaxFF and QM showing that ReaxFF can properly reproduce QM results.

Heats of formation (**Fig. 1e**.) of various crystalline $Cu_xSi_y$ species were also utilized in the force field development by using following relation:

$$\Delta E = E_{Cu_xSi_y} - xE_{Cu} - yE_{Si} \quad (3)$$

where $E_{Cu_xSi_y}$ is the total energy of the Cu-Si system, x and y are the atomic fraction of copper and silicon, respectively, and $E_{Cu}$ and $E_{Si}$ are the energies per atom for Cu and silicon, respectively. We see that ReaxFF acceptably reproduces the QM results near the equilibrium. Furthermore, we trained our force field for Cu atoms migration inside Si lattice. ReaxFF parameters were fitted against QM results which can be seen in **Fig. 1f**. We see good agreement between the ReaxFF and QM results for the Cu atom migration pathway. Overall, ReaxFF energy descriptions are in good agreement with the QM data, which establishes the capability of the force field to describe the chemistry of copper silicon interactions. Force field parameters are given in the supporting information.

## 2.3  Simulation methodology

We employed our Cu-Si force field to study Cu diffusion and dissolution inside silicon lattice. Initial geometries were constructed in three configurations: 1) placing a small 14-atom Cu crystal outside of Si lattice with 512 atoms, 2) placing a small 14-atom Cu crystal inside of Si lattice with 512 atoms, and 3) randomly dispersing individual Cu atoms at the given concentrations in a silicon lattice phase comprised of 512 atoms (**Fig. 2**). These geometries were relaxed using a conjugate gradient minimization scheme. We performed an equilibration of the box in three stages: the first stage is performed at 50 *K* for 50 *ps* to eliminate any hot spots in the initial geometry. Next, the temperature is increased from 50 K to various temperatures (i.e., 300, 500, 700, 1000, 1200, 1500, 1700, and 2000 K) at a rate of 2.5 Kps$^{-1}$ in NVT (constant volume, temperature) ensemble. Finally, in the third stage, the box is equilibrated at the target temperature for 200 ps. Temperature and pressure were regulated using the Berendsen thermostat and barostat [], respectively. These three stages were performed using the NVT ensemble with a temperature relaxation time of 100 fs at 1 atmosphere, and with a pressure relaxation time of 100 ps. Periodic boundary conditions were employed in all three directions, and a MD time step of 0.25 fs was used for all the simulations in this study.

## 3 Results and Discussion

The three configurations studied in this work are shown in **Fig. 2**. In each configuration, the initial, half-way, and ending snapshots of the Cu/Si system is shown in the first, middle, and final columns, respectively. In the first configuration, a Cu cluster consisting of 14 Cu atoms is initially placed adjacent to the silicon supercell. As indicated in **Fig. 2a**, immediately after starting the MD simulation, the Cu cluster starts to deform and diffuses into the Si supercell which is due to the high surface ratio between the Si and Cu atoms. In the second configuration, the same Cu cluster as in the previous configuration, is placed inside the Si supercell. By running the MD simulation on this structure, the Cu cluster diffuses inside the Si cluster through the interstitial sites. In the final configuration, individual Cu atoms are randomly placed within the Si interstitial sites. While the individual Cu atoms diffuse inside the Si supercell, they begin to form Cu clusters which is thoroughly investigated throughout this paper.

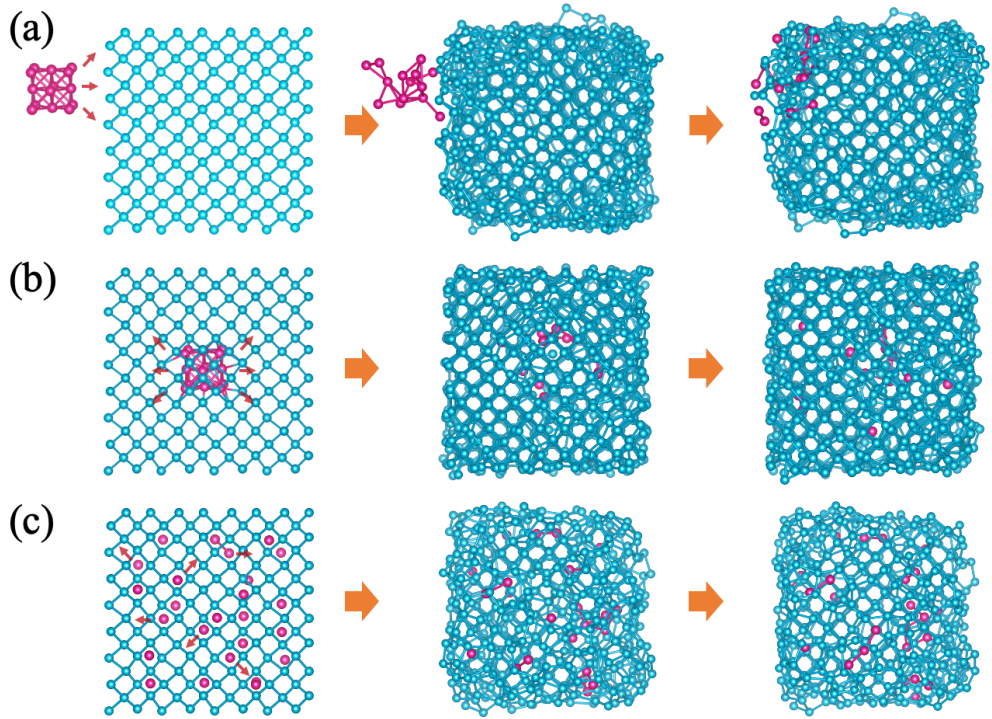

**Figure 2.** Snapshots of the Cu (pink spheres) and Si (light blue spheres) structure at the initiation, halfway, and end of the MD simulation for three configurations: (a) 14 atom Cu cluster diffuses into the 512 atom Si supercell, (b) 14 atom Cu cluster diffuses inside the 512 atom Si supercell, and (c) individual Cu atoms diffuse inside the 512 atom Si supercell.

### 3.1 Cu cluster diffusing from outside the silicon lattice

For the first scenario of our experiments, the simulations are done in canonical ensemble (NVT) at different temperatures ranging from 300K to 1000K. The Cu concentration is about 2.7 percent (14 to 512). Based on our observation, the Cu atoms start to diffuse into the Si cluster instantly at the beginning of the MD simulation. The motion of Cu atoms is traced by their mean square displacement (MSD) and it is used to calculate the diffusion coefficient. **Fig. 3** shows the MSD data for the Cu atoms migrating inside silicon lattice as a function of simulation time for four different MD temperatures (300, 500, 700, and 1000 K). The diffusion coefficient for each case is calculated by using Eq. 4 [62]:

$$D_i = \lim_{t \to 0} \frac{<[r(t)-r(0)]^2>_i}{6t} = \frac{1}{6}\frac{d(MSD)}{dt} \quad (4)$$

Where $D_i$ represents the diffusion coefficient and it is obtained from the slope of the MSD curves (**Fig. 3**). It is known from the calculations that the more the MD temperature increases; the more rapidly Cu atoms diffuse into the Si lattice. At the beginning of the simulation, the MSD curve nonlinearly increases and plateaus, which indicates that the particles are confined. In other words, the particles are stationary vibrating at the first 5ps of the simulation. The rest of the MSD plot could be considered as the diffusional behavior in which the MSD becomes more linear. The inset of **Fig. 3** shows the diffusion coefficient of Cu atoms as a function of MD temperature. The data indicates that there's an exponential relation between the diffusion coefficient and MD temperature which is in agreement with the following Arrhenius Equation[63]:

$$D(T) = D_0 \exp(-\frac{E_a}{kT}) \quad (5)$$

In the equation above, $D(T)$ is the temperature dependent diffusion coefficient, $D_0$ is the pre-exponential factor of maximum diffusion at infinite temperature, $E_a$ is the activation energy for diffusion, k is the Boltzmann constant and T is the temperature. Since the experimentally observed pre-exponential factor $D_0$ of diffusion coefficient is approximately the same value for all the 3d transition metal (TM) impurities in silicon, the large difference of diffusivity for these elements comes primarily from the difference of the diffusion barrier ($E_a$). The underlying mechanism behind this phenomenon should be directed to the electronic origin of the elements which makes the difference of the diffusion coefficient between the early 3d TM and the late 3d TM impurities in silicon almost one order of magnitude [].

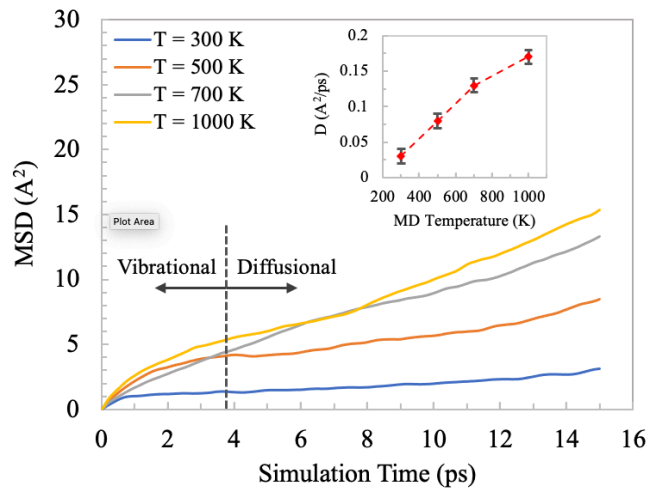

**Figure 3.** Mean squared displacement of Cu atoms as a function of simulation time running under four different temperatures from 300K to 1000K. The initial non-linear increase in the MSD belongs to the vibrational motions, while the linear section indicates the diffusional motion of Cu atoms through the interstitial sites of Si lattice. The inset shows the diffusion coefficient of copper atoms as a function of temperature.

To examine the accuracy of the ReaxFF method, and to have a better insight into the Cu diffusion, we investigated the energy barriers for diffusion of a single Cu atom through a diamond cubic super-cell of 64 Si atoms, corresponding to a 2 × 2 × 2 multiple of the 8-atom diamond cubic cell. Here, we use a bond-restraint approach to find the minimum energy pathways on the potential energy surface between two sites for Cu insertion that represents stable local energy minima. To explore the Cu migration pathways between two equilibrium neighboring sites of the Cu atom, a penalty potential function is added to the ReaxFF energy to maintain the distance between two adjacent sites using the Eq. (6):

$$E_{restraint} = f_1 \left(1 - e^{f_2(r_0 - r_{ij})^2}\right) \quad (6)$$

In this equation, $f_1$ and $f_2$ are constants; during each MD step, the value of the distance between two sites $r_{ij}$ is modified to sample the pathway between two local minima separated by the initial distance $r_0$. For the case of a single Cu atom in c-Si, the thermodynamically favorable site for Cu atom insertion is the tetrahedral ($T_d$) interstitial. The Cu diffusion path consists of the Cu migration between two adjacent $T_d$ interstitial sites passing through the hexagonal (Hex) site. A comparison of the energy landscape along the diffusion pathway obtained by DFT and ReaxFF are shown in **Fig. 4**. The initial, saddle-point, and final state of Cu atom are also indicated throughout this transition. The energy barrier height obtained from our MD simulation is about 1.1 eV, which is in reasonable agreement with the DFT result (1.5 eV).

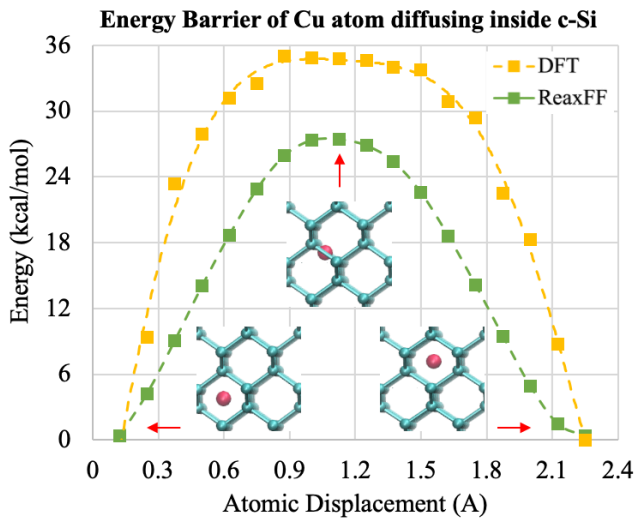

**Figure 4.** Energy landscape for diffusion of an individual Cu atom in c-Si from one equilibrium $T_d$ site to an adjacent one through the hexagonal configuration calculated by DFT (yellow) and ReaxFF (green). The light blue spheres represent Si atoms, the pink spheres represent the Cu atom in its initial, saddle-point and target (final) configuration.

### 4.1 Copper cluster diffusing form inside the silicon lattice

In the following section, we studied the diffusional behavior of copper atoms inside bulk silicon. The initial 14-atom Cu cluster is placed within the same silicon cluster similar to the last section (**Fig. 2 b**). MD simulations are performed at temperatures ranging from 1000 K to 1700 K in NVT ensemble. To trace the motion of the system, MSD data for Cu(a) and Si(b) atoms are plotted as a function of simulation time for four different temperatures (1000, 1200, 1500, and 1700 K) in **Fig. 5**. The MSD plot shows that Cu diffusion at lower temperatures is not considerable. This means that higher temperatures (above 1000K) are required to trigger the Cu diffusion. On the other hand, silicon atoms do not show any diffusion in comparison with the Cu atoms indicating that Si lattice keeps its crystalline structure throughout the simulation.

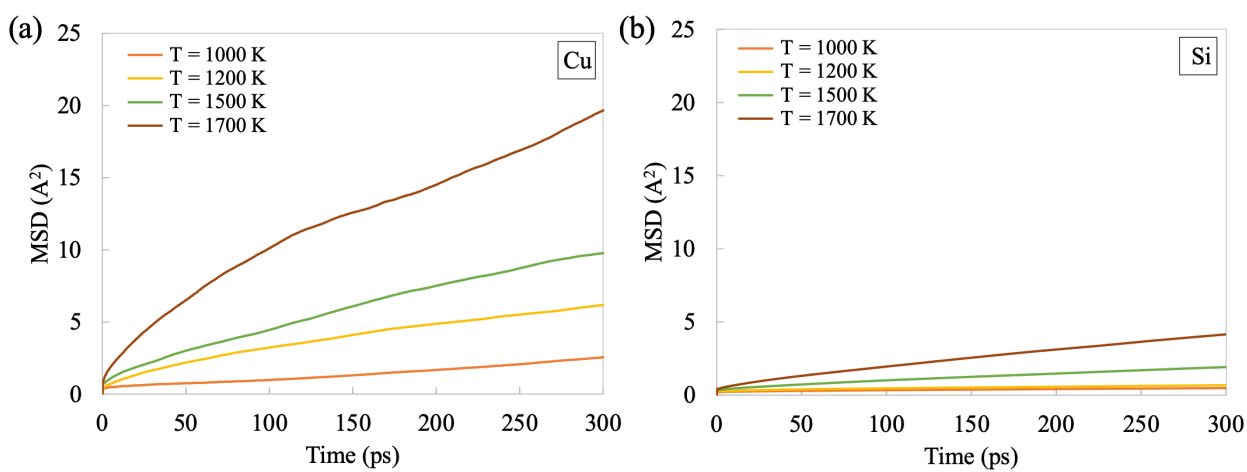

**Figure 5.** Mean squared displacement of Cu (a) and Si (b) atoms as a function of simulation time at different temperatures ranging between 1000K and 1700K.

In line with the MSD analysis, we looked at the Radial Distribution Function (RDF) of Cu atoms inside Si. The integration of the RDF for Cu atoms as a function of radius for the MD simulation at 1500 K is shown in **Fig. 6**. Each line of data indicates the integral on a certain time during the MD simulation moving from the start (darker) to the end (lighter). Since Int(gr) illustrates the coordination number of the atoms, it can provide an insight of the number of neighboring atoms as a function of time. At the beginning of simulation (the darkest line), Cu atoms are keeping their crystalline form. Over time, Cu cluster loses its crystalline structure and diffuses into the Si. In order to look at the number of neighboring Cu atoms within the radius of 3.2 A, we can simply draw a vertical line in **Fig. 6a** at the radius of 3.2 A and plot the integral of RDF as a function of simulation time. As shown in **Fig. 6b**, the number of neighboring atoms start to decrease as the Cu atoms diffuse into the Si throughout the simulation. The inset of **Fig. 6b** shows the RDF of Cu-Cu, Si-Si, and Cu-Si atom pairs as a function of radius. It is also noteworthy to mention that g(r) peaks approximately at 2.3 and 2.7 A for Si and Cu atoms, respectively. This observation specifies the Cu-Cu and Si-Si bond lengths which strongly agree with the experimental data [] supporting the quality of our Cu/Si force field.

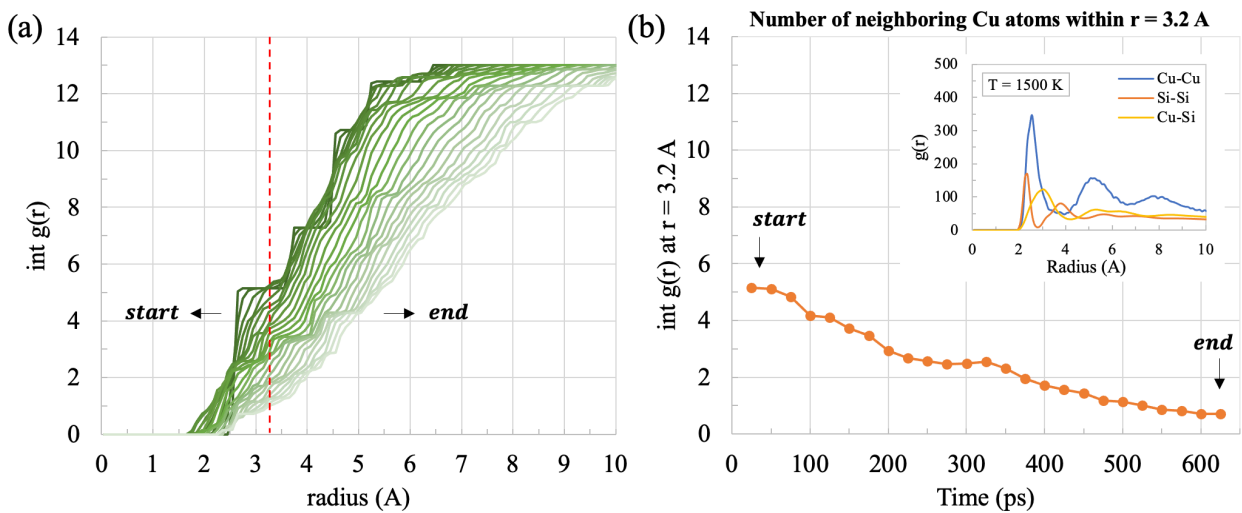

**Figure 6.** (a) Integral of radial distribution function for Cu atoms as a function of radius for the MD simulation at 1500K. (b) Number of neighboring copper atoms within 3.2A as a function of simulation time. The inset shows the radial distribution function for Cu-Cu, Si-Si, and Cu-Si atom pairs as a function of radius.

### 4.2 Copper atoms diffusing from inside the silicon lattice

For the final configuration of Cu/Si system, we looked at individual Cu atoms randomly placed inside Si sites within the Si lattice. The crystallization of Cu atoms inside Si crystal is investigated by running NVT molecular dynamic simulations at different temperatures and Cu concentrations. The snapshots of the Cu migration inside Si structure during simulation under three temperature regimes (500K, 1250K, 2000K) with four initial levels of copper concentration (2, 5, 10, and 16%) are shown in **Fig. 7**. For these experiments, the MD temperature is fixed at 500, 750, 1000, 1250, 1500, 1750, and 2000 K, and the Cu concentration is set to 2, 3, 5, 10, 16, 30, and 49 percent. The simulation timestep is set to 0.25 fs making the cumulative simulation time of 725 ps. An interesting behavior of Cu atoms observed during the simulations is that they start to approach to each other and form bonds with the surrounding copper atoms which plays the main role in the formation of copper crystal during the simulation. As indicated in **Fig. 7**, Cu crystallization is strongly related to the Cu concentration. Bigger crystals and chain structures of Cu atoms form by increasing the Cu concentration. In addition, higher temperature helps with Cu crystallization since the Cu atoms need a starting momentum to overcome the energy barrier and diffuse through the silicon lattice. However, it can be observed from **Fig. 7** that Cu crystallization is decreased at the highest temperature. This is due to the fact that very high temperatures make the kinetic energy of Cu atoms high enough to prevent them to stay at one place and maintain the existing bonds with the surrounding Cu atoms. In addition, high temperature lead to thermal stresses on the system and breaking the formed bond in Cu cluster. Thermal stress is a type of stress that remains in the matrix after clustering.

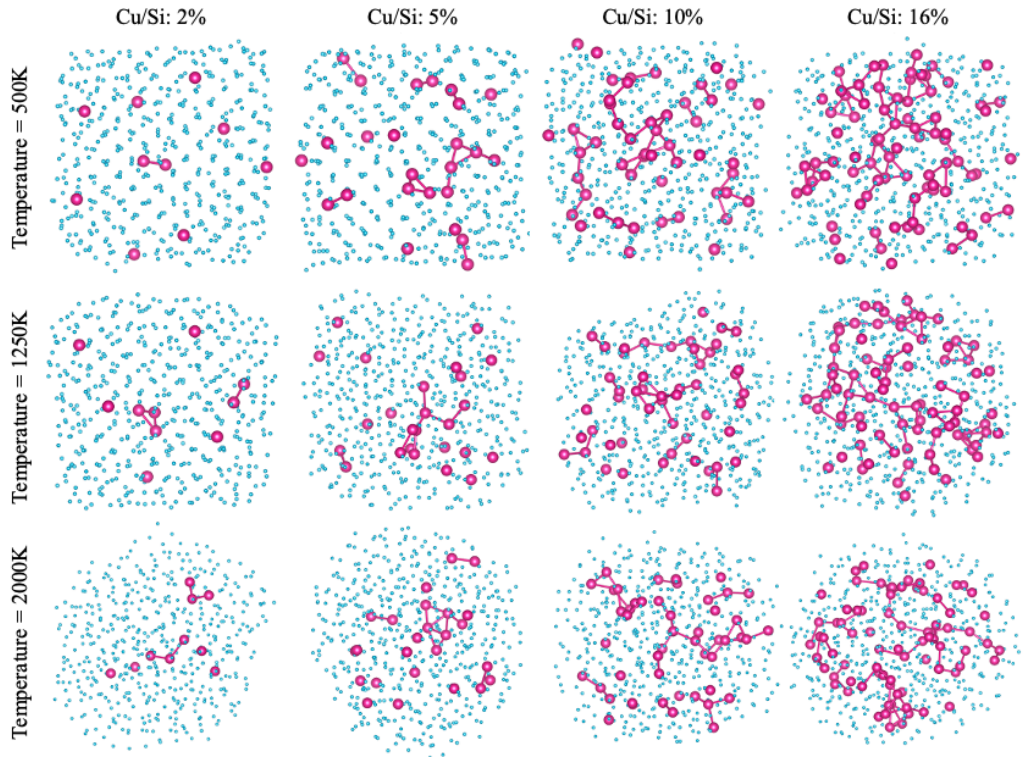

**Figure 7.** Snapshots of the Cu (pink spheres) and Si (light blue spheres) structure while running the MD simulation under three temperature regimes (500K, 1250K, 2000K) with four initial levels of copper concentration (2, 5, 10, and 16%).

### 4.3 Copper crystallization

Due to its direct impact on device performance, an important feature of the Cu clusters formed during the simulation is their structural properties. The overall electrical conductivity of the system is significantly altered by the existence of chain, loop, or diamond shape Cu clusters. Therefore, we analyzed the structure of the Cu/Si configurations by using connected-component clustering (CCC) algorithm []. Using CCC, for each simulation at a certain time and a specific Cu concentration, we can import the Cartesian coordinates of Cu atoms inside the structure, calculates the Euclidian distance for each copper atom pairs, and classify them into separate clusters as a function of simulation time. This classification is based on the bonds created between individual copper atoms which can be extended to any other element of interest. Since the optimized bond length between Cu-Cu atom pair is approximately 2.7 A, the maximum bond length used for this classification is 3 A. Following this approach, we can obtain useful structural information about our system such as the size and geometry of the Cu clusters during the simulation, the number of Cu clusters within the Si crystal, RDF of Cu atoms, and the number of 3-members Cu rings within the structure. 3-members Cu rings are the simplest form of Cu clusters that can easily attract an additional Cu atom and form a 4-members Cu diamond.

The average size of the largest Cu cluster as a function Cu concentration for the simulation at 1000 K is shown in **Fig. 8a**. As indicated by the plot, Cu concentration has direct impact on Cu crystallization. **Fig. 8b** shows the time-trace of the average size of the largest Cu cluster as for the simulations at 1000 and 2000 K. As indicated in the figure, at 2000 K, Cu atoms show longer and recurrent movements inside the Si cluster which is due to their higher Kinetic energy. However, at 1000 K, Cu atoms show limited movements which leads to less change in the size of the largest Cu cluster.

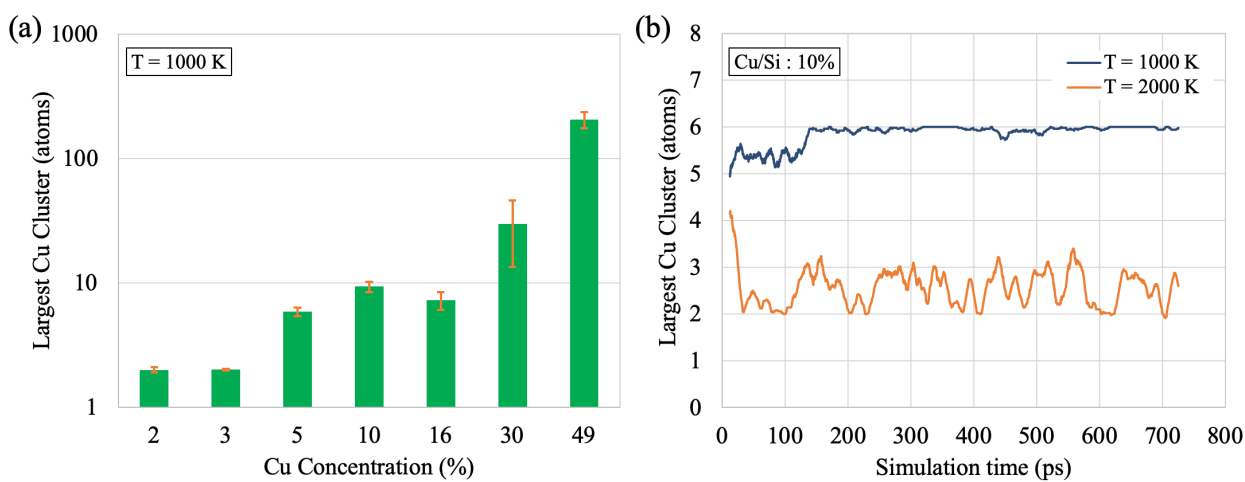

**Figure 8.** (a) Average number of copper clusters formed during the MD simulation in NVT ensemble at 1000 K as a function of initial Cu concentration. (b) Integral of radial distribution function for copper atoms as a function of radius at 1000 K.

In addition, CCC also generates the number of 3-atom Cu rings throughout the simulation (**Fig. 9a**). Based on this plot, the number of Cu rings is directly related to the Cu concentration. As indicated in **Fig. 9a**, Cu concentrations of 3% and below don't produce any Cu rings. However, there is an exponential relation between the number of Cu rings and the Cu concentrations above 5%. Moreover, we claim that the number of Cu ring has reverse relation with the MD temperature.

Draft for Submission to XXX

**Fig. 9b** shows the average number of 3-members Cu rings as a function of temperature. We observed that at lower temperatures more Cu rings were formed during the simulation which is in agreement with our previous argument. However, at higher temperatures, due to the higher kinetic energy and rapid movements of Cu atoms through the Si sites, less number of 3-members Cu rings are observed.

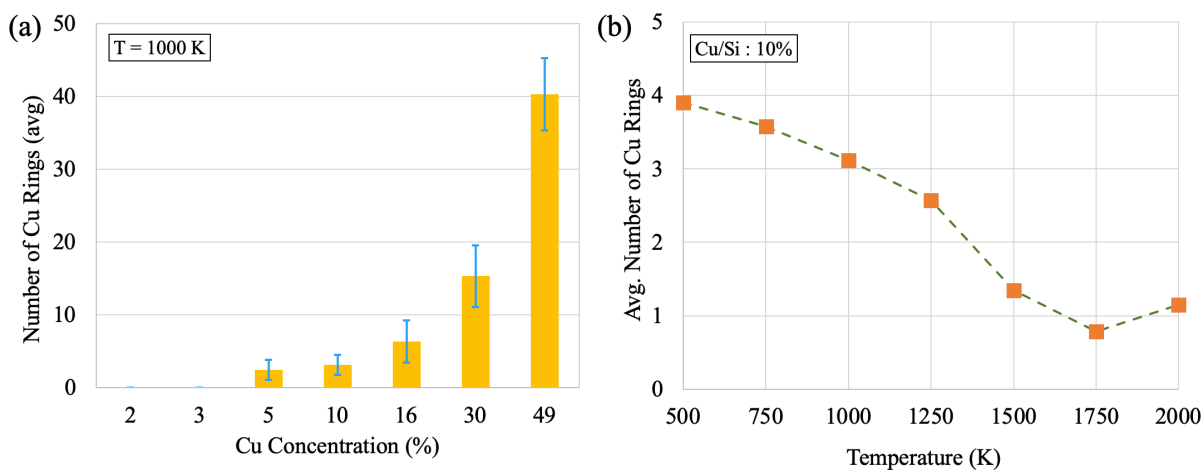

**Figure 9.** (a) Average number of Cu 3-atom rings formed during the MD simulation in NVT ensemble at 1000 K as a function of initial Cu concentration. (b) Integral of radial distribution function for copper atoms as a function of radius at 1000 K.

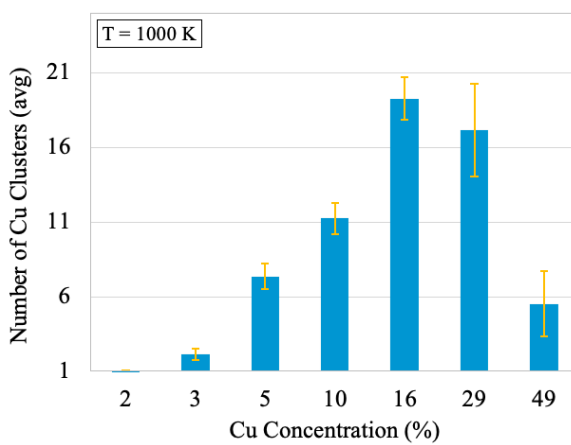

**Figure 10.** Average number of Cu clusters formed during the MD simulation in NVT ensemble at 1000 K as a function of Cu concentration.

Another interesting observation from the analyzed results of CCC algorithm is the number of Cu clusters during the simulation as a function of Cu concentration. **Fig. 10** reveals that during the simulation at 1250 K, no Cu clusters of any form is observed at lower Cu concentrations. On the other hand, increased Cu contaminations will usually lead to the formation of a single large Cu cluster. Based on this observation, we can argue that with around 16% initial Cu concentration, the maximum number of Cu clusters can be observed during the simulation at 1000 K. **Fig. 11** shows the integral of RDF for the Cu-Cu atom pairs inside the Si crystal for different initial Cu concentrations by running the simulation at 1000 K. As expected, the number of neighboring Cu atoms increases by increasing the Cu concentration. An interesting observation from **Fig. 11** is that there is a slight gap between the number of neighboring atoms for the Cu concentrations of (3%-under) and (5%-above). This observation indicates that concentration of 5% and above are more likely to form Cu clusters during the simulation.

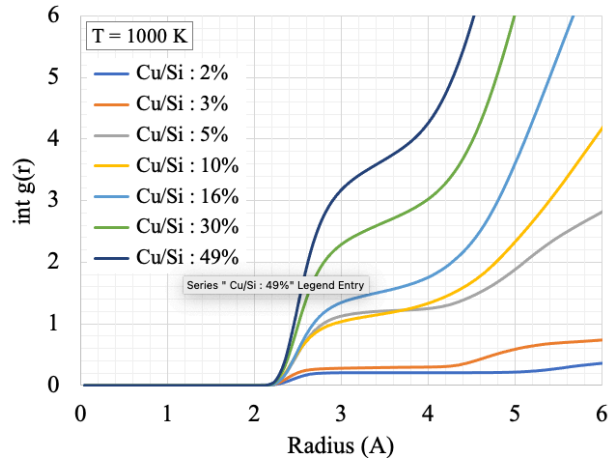

**Figure 11.** Integral of radial distribution function for Cu atoms as a function of radius at 1000 K.

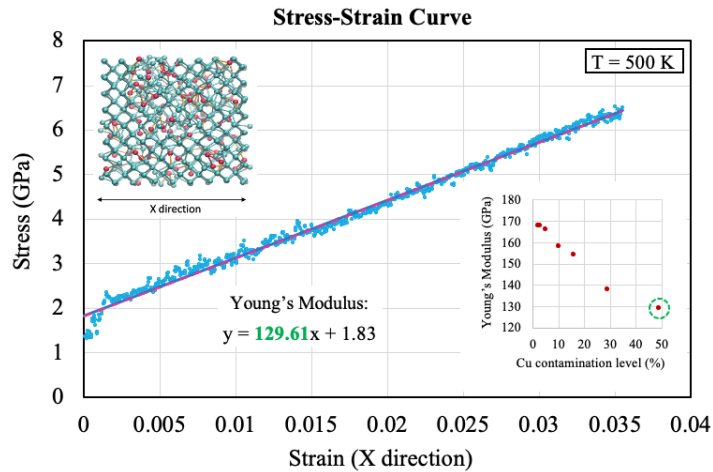

**Figure 12.** Tensile stress-strain curves at strain rate of $2.5 \times 10^{-7}$ A/iteration during the MD simulation for the sample with 29% Cu concentration inside Si lattice at T = 500 K. The Young's modulus is calculated using linear fit to the stress-strain curve. The inset shows the Young's modulus as a function of Cu concentration.

### 4.4 Mechanical properties

To characterize the mechanical behavior of the Cu/Si compounds, we performed MD simulations at room temperature to obtain the stress-strain relationship under uniaxial tensile loading. We considered seven different samples (2, 3, 5, 10, 16, 29 and 49% Cu concentration) to investigate the effect of Cu contamination on the mechanical properties of the system. **Fig. 12** shows tensile stress-strain curves at strain rate of $2.5\times10^{-7}$ A/iteration during the 125 ps simulation time for the sample with 29% Cu concentration inside Si lattice. Similar trends in the stress-strain curve at other Cu concentration have been observed in Cu/Si lattice deformation simulations. In general, stress-strain curves exhibit an initial linear region followed by a nonlinear portion. The linear portion of the uniaxial stress-strain curve corresponds to elastic deformation, and the gradient of this part is Young's modulus (YM). In this study, YM was calculated using linear regression on the initial linear portion. The calculated result for YM for various compositions of Cu has been shown in table 1 and they are plotted in the inset of **Fig. 12**. The results demonstrate that YM will drop with increasing Cu concentration which is in good agreement with experimental data []. Special attention should be drawn to the decrease in elastic modulus while increasing the level of impurity concentration. In this phenomenon spontaneous microcracking of the Si matrix occurs during stress-strain test. Microcrack propagation in the presence of an external load applied on the Si lattice has been shown in **Fig. 13**. A small cluster of Cu is formed where the cracking initiates, and as a result, the nearby Si-Si bonds brake over the course of our simulation. Inclusions can also be fracture initiators due to the thermal residual stresses introduced during processing, and the intensification of stress in the surroundings of the inclusions that occurs when an external load is applied.

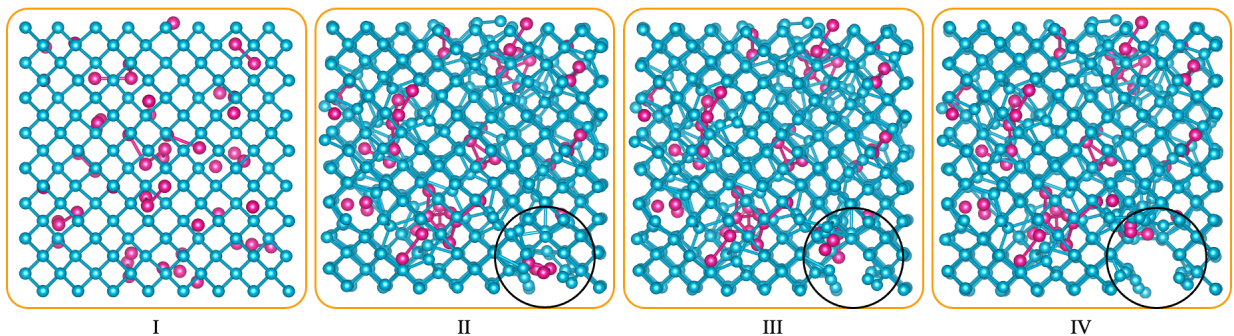

I  II  III  IV

**Figure 13.** Snapshots of the Cu (pink spheres) and Si (light blue spheres) structure while running the MD simulation indicating the crack propagation when Cu impurities are present in Si lattice under a stress load.

Diffusion of copper atoms into the silica layer is one of the important problems for ultra-large-scale integration (ULSI) metallization. Although copper concentrations studied in this work are beyond practical values, the findings of this study can be further accompanied with Machine Learning implementations to predict certain electromechanical properties of Cu/Si compounds. For instance, numerous samples of various environmental conditions, structural parameters, initial concentrations, and electrical conductivity of the system calculated by Tight Binding DFT (DFTB) techniques can be given to the AI as an input, and in return, AI learns to predict the electrical conductivity given the initial status of the system. DFTB methods offer handful of advantages including the ability to analyze larger systems and accessing longer timescales.

## 5 Conclusion

In this study, we developed the Cu/Si force field parameters to understand the atomic interactions between Cu and silicon by studying the diffusion behavior of copper impurities placed inside silicon lattice. We compared the migration barriers of Cu in amorphous Si obtained by ReaxFF with those obtained by DFT calculations to validate the Cu/Si ReaxFF force field. In addition, we introduced a new computational study to atomistically model various configurations copper atoms incorporate as a Cu cluster inside the silicon lattice. In this part, we studied the dependence of Cu crystallization on Cu contamination level as well as environmental conditions such as the MD temperature. We found that Cu crystallization has a direct relation with Cu concentration. Lastly, mechanical properties of Cu/Si compounds have been investigated. Our results demonstrate the decisive role of stresses in the migration of the Cu inside pristine Si, which leads to micro cracks propagation.

## 6 Associated Content

**Supporting Information**
The Supporting Information is available free of charge on the XXX Publications website at DOI: 10.1021/acs.jpcc.8b05852.

## 7 Author Information

**Corresponding Author**
E-mail: acv13@engr.psu.edu. Tel: +1-814-863-6277.**Corresponding Author**
E-mail: acv13@engr.psu.edu. Tel: +1-814-863-6277.

# 8 Acknowledgments

This work was supported by NSF DMR grant #1609107. Computations for this research were performed on the Pennsylvania State University's Institute for Cyber Science Advanced Cyber Infrastructure (ICS-ACI).

Draft for Submission to XXX[9]   D. V. Talapin, J.S. Lee, M. V. Kovalenko, E. V. Shevchenko, Prospects of colloidal nanocrystals for electronic and optoelectronic applications, Chem. Rev. (2010). https://doi.org/10.1021/cr900137k.

[10]  C. Lee, J. Meteer, V. Narayanan, E.C. Kan, Self-assembly of metal nanocrystals on ultrathin oxide for nonvolatile memory applications, J. Electron. Mater. (2005). https://doi.org/10.1007/s11664-005-0172-8.

[11]  Y.-L. Cheng, T.-J. Chiu, B.-J. Wei, H.-J. Wang, J. Wu, Y.-L. Wang, Effect of copper barrier dielectric deposition process on characterization of copper interconnect, J. Vac. Sci. Technol. B, Nanotechnol. Microelectron. Mater. Process. Meas. Phenom. (2010). https://doi.org/10.1116/1.3425631.

[12]  J. Čechal, J. Polčák, M. Kolíbal, P. Bábor, T. Šikola, Formation of copper islands on a native SiO 2 surface at elevated temperatures, Appl. Surf. Sci. (2010). https://doi.org/10.1016/j.apsusc.2009.12.168.

[13]  S. V. Kershaw, A.S. Susha, A.L. Rogach, Narrow bandgap colloidal metal chalcogenide quantum dots: Synthetic methods, heterostructures, assemblies, electronic and infrared optical properties, Chem. Soc. Rev. (2013). https://doi.org/10.1039/c2cs35331h.

[14]  D.M. Guzman, N. Onofrio, A. Strachan, Stability and migration of small copper clusters in amorphous dielectrics, J. Appl. Phys. 117 (2015) 1–9. https://doi.org/10.1063/1.4921059.

[15]  J. Lindroos, D.P. Fenning, D.J. Backlund, E. Verlage, A. Gorgulla, S.K. Estreicher, H. Savin, T. Buonassisi, Nickel: A very fast diffuser in silicon, J. Appl. Phys. (2013). https://doi.org/10.1063/1.4807799.

[16]  A.A. Istratov, C. Flink, H. Hieslmair, E.R. Weber, T. Heiser, Intrinsic diffusion coefficient of interstitial copper in silicon, Phys. Rev. Lett. (1998). https://doi.org/10.1103/PhysRevLett.81.1243.

[17]  N. Nejatishahidein, E.E. Borujeni, D.J. Roush, A.L. Zydney, Effectiveness of host cell protein removal using depth filtration with a filter containing diatomaceous earth, Biotechnol. Prog. (2020). https://doi.org/10.1002/btpr.3028.

[18]  K.A. Roshan, Z. Tang, W. Guan, High fidelity moving Z-score based controlled breakdown fabrication of solid-state nanopore, Nanotechnology. (2019). https://doi.org/10.1088/1361-6528/aaf48e.

[19]  K.A. Roshan, Z. Tang, W. Guan, False Negative and False Positive Free Nanopore
24

Draft for Submission to XXX071312-121610.

[49]  A.C.T. Van Duin, A. Strachan, S. Stewman, Q. Zhang, X. Xu, W.A. Goddard, ReaxFFSiO reactive force field for silicon and silicon oxide systems, J. Phys. Chem. A. 107 (2003) 3803–3811.

[50]  K.D. Nielson, A.C.T. Van Duin, J. Oxgaard, W.Q. Deng, W.A. Goddard, Development of the ReaxFF reactive force field for describing transition metal catalyzed reactions, with application to the initial stages of the catalytic formation of carbon nanotubes, J. Phys. Chem. A. (2005). https://doi.org/10.1021/jp046244d.

[51]  T.P. Senftle, S. Hong, M.M. Islam, S.B. Kylasa, Y. Zheng, Y.K. Shin, C. Junkermeier, R. Engel-Herbert, M.J. Janik, H.M. Aktulga, T. Verstraelen, A. Grama, A.C.T. Van Duin, The ReaxFF reactive force-field: Development, applications and future directions, Npj Comput. Mater. 2 (2016) 1–14.

[52]  M. V. Fedkin, Y.K. Shin, N. Dasgupta, J. Yeon, W. Zhang, D. Van Duin, A.C.T. Van Duin, K. Mori, A. Fujiwara, M. Machida, H. Nakamura, M. Okumura, Development of the ReaxFF Methodology for Electrolyte-Water Systems, J. Phys. Chem. A. (2019). https://doi.org/10.1021/acs.jpca.8b10453.

[53]  H. Kwon, B.D. Etz, B.D. Etz, M.J. Montgomery, R. Messerly, S. Shabnam, S. Vyas, A.C.T. Van Duin, C.S. McEnally, L.D. Pfefferle, S. Kim, Y. Xuan, Reactive Molecular Dynamics Simulations and Quantum Chemistry Calculations to Investigate Soot-Relevant Reaction Pathways for Hexylamine Isomers, J. Phys. Chem. A. (2020). https://doi.org/10.1021/acs.jpca.0c03355.

[54]  B. Damirchi, M. Radue, K. Kanhaiya, H. Heinz, G.M. Odegard, A.C.T. Van Duin, ReaxFF Reactive Force Field Study of Polymerization of a Polymer Matrix in a Carbon Nanotube-Composite System, J. Phys. Chem. C. (2020). https://doi.org/10.1021/acs.jpcc.0c03509.

[55]  J. Sun, J.W. Furness, Y. Zhang, Density functional theory, in: Math. Phys. Theor. Chem., 2018. https://doi.org/10.1016/B978-0-12-813651-5.00004-8.

[56]  W. Kohn, L.J. Sham, Self-consistent equations including exchange and correlation effects, Phys. Rev. (1965). https://doi.org/10.1103/PhysRev.140.A1133.

[57]  N.A. Pike, O.M. Løvvik, Calculation of the anisotropic coefficients of thermal expansion: A first-principles approach, Comput. Mater. Sci. (2019). https://doi.org/10.1016/j.commatsci.2019.05.045.28

Draft for Submission to XXX